\documentclass[]{aipproc}

\layoutstyle{6x9}

\SetInternalRegister\hbadness{8000} 
\newcommand{\bea}{\begin{eqnarray}}
\newcommand{\eea}{\end{eqnarray}}
\newcommand{\beqa}{\begin{eqnarray}}
\newcommand{\eeqa}{\end{eqnarray}}
\newcommand{\beq}{\begin{equation}}
\newcommand{\eeq}{\end{equation}}

\newcommand{\apks}{\mbox{$a_{\rm CP}(B\to \psi K_S)$}}
\newcommand{\apkl}{\mbox{$a_{\rm CP}(B\to \psi K_L)$}}

\newcommand{\ltap}{\stackrel{<}{_\sim}}
\newcommand{\gtap}{\stackrel{>}{_\sim}}

\begin{document}

\title{Beyond the Standard Model in $B$ Decays: Three Topics\footnote{Invited talk at the
9th International Symposium on Heavy Flavor Physics, Caltech, Pasadena, Sept. 10-13, 2001}}


\author{Alexander L. Kagan 
\thanks{Work supported by the Department of Energy, 
Grant No. DOE DE FG02-84ER-40153} }
{address={Physics Department, University of Cincinnati, Cincinnati OH 45221, USA}}

\copyrightyear  {2001}

\begin{abstract}
Three new results are discussed:
(a)  A non-vanishing amplitude for the `wrong sign{'} kaon decay
$B \to J/\Psi \overline{K}$ or its CP conjugate is shown to be a necessary condition for 
obtaining different CP asymmetries in $B \to J/\Psi K_{S,L} $. A significant effect would require
a scale of new physics far below the weak scale, all but ruling out this possibility.   
(b) The leading isospin breaking contributions to the 
$B \to K^* \gamma $ decay amplitudes can be calculated in QCD factorization,
providing a sensitive probe of the penguin sector of the effective weak Hamiltonian.
New physics models which reverse the predicted $10-20\% $ 
Standard Model amplitude hierarchy 
could be ruled out with more precise data. 
(c) A slowly falling $g^* g \, \eta^\prime $ form factor
can be ruled out using the $\eta^\prime $ spectrum obtained by ARGUS at 
the $\Upsilon (1S) $.  The decay $b \to s g \eta^\prime $ is therefore
highly suppressed and the origin of the anomalously large
$B \rightarrow \eta^\prime X_s $ rate 
remains unknown, perhaps requiring the intervention 
of New Physics.

\end{abstract}
\date{\today}

\maketitle

\section{Introduction}

Three recent developments with potential implications for new physics and $B$ decays are discussed.
Recently, the BaBaR and Belle collaborations presented new measurements 
of the time-dependent CP asymmetries in $B \to J/\Psi K_S $ and $B \to J/\Psi K_L $~
\cite{belleCP,babarCP}.  In the Standard Model the two CP asymmetries are predicted to have the same 
magnitude but opposite sign.  Although the measured values are equal within errors
the CP asymmetry in the $B \to J/\Psi K_L $ channel is somewhat larger, which raises the question
of whether $|\apkl| \ne |\apks|$ is possible.  The conditions which must be fulfilled
for this to happen are explained below~\cite{psikskl}.  A model-independent analyis
implies that the associated scale of new physics interactons would have to lie at a 
prohibitively low scale of a few GeV in order to 
obtain a significant effect.

Measurements of the exclusive $B\to K^*\gamma$ branching ratios 
have been reported by the CLEO, Belle and BaBar Collaborations, with the 
results (averaged over CP-conjugate modes):
\begin{eqnarray}
   10^5\,\mbox{Br}(\bar B^0\to\bar K^{*0}\gamma) 
   &=& \cases{ 4.55_{\,-\,0.68}^{\,+\,0.72}\pm 0.34 &
                ~\protect\cite{Coan:2000kh} \cr
               4.96\pm 0.67\pm 0.45 & ~\protect\cite{Ushiroda:2001sb} \cr
               4.23\pm 0.40\pm 0.22 & ~\protect\cite{BaBarKsgamma} }
    \nonumber\\
   10^5\,\mbox{Br}(B^-\to\bar K^{*-}\gamma) 
   &=& \cases{ 3.76_{\,-\,0.83}^{\,+\,0.89}\pm 0.28 &
                ~\protect\cite{Coan:2000kh} \cr
               3.89\pm 0.93\pm 0.41 & ~\protect\cite{Ushiroda:2001sb} \cr
               3.83\pm 0.62\pm 0.22 & ~\protect\cite{BaBarKsgamma} }
    \nonumber
\end{eqnarray}
The average branching ratios for the two modes are 
$(4.44\pm 0.35)\cdot 10^{-5}$ and $(3.82\pm 0.47)\cdot 10^{-5}$. When 
corrected for the difference in the $B$-meson lifetimes, 
$\tau_{B^-}/\tau_{\bar B^0}=1.068\pm 0.016$~\cite{Blife}, these results 
imply
\beq
   \Delta_{0-}\equiv\frac{\Gamma(\bar B^0\to\bar K^{*0}\gamma)
                          -\Gamma(B^-\to\bar K^{*-}\gamma)}
                         {\Gamma(\bar B^0\to\bar K^{*0}\gamma)
                          +\Gamma(B^-\to\bar K^{*-}\gamma)}
   = 0.11\pm 0.07 \,.
\label{eq:isospinavg}
\eeq
Although there is no significant deviation of this quantity from zero, 
the fact that all three experiments see a tendency for a larger neutral 
decay rate raises the question of whether the Standard Model can account
for isospin-breaking effects of order 10\% in the decay amplitudes.

Recently it has been shown that in the heavy-quark limit the 
decay amplitudes for these processes can be calculated in a 
model-independent way using a QCD factorization approach~ 
\cite{Bosch:2001gv,Beneke:2001at}, which is similar to the scheme 
developed for the analysis of nonleptonic two-body decays of $B$ mesons~\cite{BBNS}. 
To leading order in $\Lambda/m_b$ one finds that the amplitudes for the 
decays $\bar B^0\to\bar K^{*0}\gamma$ and $B^-\to K^{*-}\gamma$ 
coincide.  Here we report on subsequent work~\cite{kaganneubert}, in which  
the QCD factorization approach was 
used to estimate the leading isospin-breaking contributions for the 
$B\to K^*\gamma$ decay amplitudes in the Standard Model.  These are due 
to annihilation graphs which enter at order $\Lambda/m_b$.  
Because of their relation to matrix
elements of penguin operators, we will see that isospin-breaking effects in $B\to K^*\gamma$ decays 
are sensitive
probes of physics beyond the Standard Model.

Finally, the CLEO collaboration and more recently, as we heard at this workshop, 
the BaBaR collaboration have
measured very large rates for fast $\eta^\prime $ production in $B \to \eta^\prime X_s $ decays:
\begin{equation}
   {\cal BR}(B \to \eta^\prime X_s )_{p_{\eta^\prime} > 2~{\rm GeV}} = \cases{
   6.2\pm 1.6\pm 1.3^{+0}_{-1.5}  \times 10^{-4}  \,; & {\rm CLEO}~\protect\cite{CLEOetap},
\vspace{0.1cm} \cr
        6.8^{+.7}_{-1.0} \pm 1^{+0}_{-.5} \times 10^{-4} \,; & {\rm BaBar}~\protect\cite{BaBaretap}. \cr}
\label{eq:etapBR}
\end{equation}
The experimental cut on 
$p_{\eta^\prime}$ is beyond the
kinematic limit for most $b \to c$ decays. 
A majority of the events lie at large recoil mass, consistent with
a three-body or higher multiplicity decay.
The $\eta^\prime$ yield from the $b \to c$ component is dominated by intermediate charmonia decays,
which contribute only $\approx 1.1 \times 10^{-4}$~\cite{atwoodsoni}
to the branching ratio. Charmless $\eta^\prime$ production proceeding via the quark content of the
$\eta^\prime$ has been estimated using factorization~\cite{pakvasa,kaganpetrov},
giving a contribution to the branching ratio of $\approx 1 \times 10^{-4}$ with
a quasi two-body recoil spectrum that is peaked at low energies in conflict
with the observed spectrum.

The surprisingly large $\eta^\prime$ yield led Atwood and Soni~\cite{atwoodsoni} to propose
that it is associated with the gluonic content of the
$\eta^\prime $ via the subprocess $b \to s (g^* \to \eta^\prime  g)$.  
Making the key
assumption that the form factor remains constant
up to $q^2 \sim m_b^2 $, where $q$ is the virtual gluon's momentum, they showed
that the large $\eta^\prime$ yield could easily be reproduced
in the Standard Model.
Moreover, they observed that the three-body decay leads to an $\eta^\prime$ recoil spectrum that is
consistent with observation. 
Hou and Tseng~\cite{houtseng} argued that the factor of $\alpha_s$
implicit in $H$ should be 
evaluated at the scale of momentum transfer through the $ g^* g \,\eta^\prime $ vertex.
However this would only introduce a mild logarithmic suppression of the form factor versus $q^2$
and therefore still lead to a large $\eta^\prime$ yield.

Assymptotically, perturbative QCD (pQCD) predicts that the leading form factor contributions should 
fall like $1/q^2$.  The question is in what region of $q^2$ does this behaviour set
in?  We will see~\cite{chenkagan} that the $\eta^\prime $ spectrum measured in $\Upsilon (1S) \to
\eta^\prime X$ decays by the ARGUS Collaboration~\cite{ARGUS} rules out form factors 
which fall slowly in the range $q^2 \le m_b^2 $.  However, a rapidly falling form factor~\cite{kaganpetrov}
representative of pQCD predictions~\cite{muta,aliformfactor} is consistent with the data.
The corresponding $b \to s g
\eta^\prime$ branching ratio in the Standard Model is about a factor 
of 20 smaller than the measured values in Eq.~(\ref{eq:etapBR}).

\section{Can the CP Asymmetries in $B \to J/\Psi K_{S,L} $ Be Different?}

The relevant quantities are~\cite{nir}
\beq
\lambda_{S,L} \equiv {q_B \over p_B } 
{\bar A_{S,L} \over A_{S,L}}\,,
\eeq
where
\beq
\bar A_{S,L} \equiv A(\bar B \to \psi K_{S,L}), \qquad
A_{S,L} \equiv A(B \to \psi K_{S,L})\,.
\nonumber \eeq
The neutral B and K meson mass eigenstates are defined in the usual way in terms of 
flavor eigenstates,
\beq
|B_{L,H}\rangle = p_B |B \rangle \pm q_B |\bar B \rangle, \qquad 
|K_{S,L}\rangle = p_K|K \rangle \pm q_K |\bar K \rangle \,.
\nonumber \eeq
The time-dependent CP asymmetries are given by 
\beq
a_{\rm CP}(B\to \psi K_{S,L})  = - 2 {{\rm{Im}}\lambda_{S,L} \over 
1 + |\lambda_{S,L}|^2 } \sin\Delta m_B t + {1 - |\lambda_{S,L}|^2 \over 1 + |\lambda_{S,L}|^2 }
\cos\Delta m_B t \,.\nonumber
\eeq
In the limit of no direct CP violation ($|\lambda_{S,L} | = 1$)
the asymmetries reduce to the simple form $ a_{\rm CP}(B\to \psi K_{S,L}) = 
- {\rm{Im}}\lambda_{S,L} \sin\Delta m_B t$,
as in the Standard
Model.

We need to rewrite $\lambda_{S,L}$ in terms of the decay amplitudes into
kaon flavor eigenstates,
\beqa
\bar A_K \equiv A(\bar{B} \to \psi K), \qquad &&
\bar A_{\bar K} \equiv A(\bar{B} \to \psi \bar K), \\ 
A_K \equiv A(B \to \psi K), \qquad &&
A_{\bar K} \equiv A(B \to \psi \bar K )\,. \label{eq:kaonamps}
\eeqa
Allowing for the possibility that the `wrong-sign{'} kaon amplitudes
$\bar A_K $ and $A_{\bar K} $ receive new physics contributions 
(they are negligible in the Standard
Model), one obtains
\beq
\lambda_{S,L} = \pm \lambda
\left( {1\pm a \over 1\pm b}\right) \,,
\nonumber
\eeq
where $\lambda \equiv q_B q_K \bar{A}_{\bar K} / p_B  p_K  A_K $, and
$a$ and $b$ are proportional to ratios of wrong-sign to right-sign kaon amplitudes
\beq
a \equiv {p_K \bar{A}_K \over q_K \bar{A}_{\bar K}}\,, \qquad
b \equiv {q_K A_{\bar K} \over  p_K A_K}\,.
\nonumber
\eeq
In the Standard Model and more generally, 
in any model in which the wrong-sign amplitudes 
are negligibly small ($a=b=0$), this reduces to $\lambda_{S,L} =  \pm \lambda $
so that ${\rm{Im} } (\lambda_S + \lambda_L ) = 0 $.  The $\sin \Delta m_b t $ term 
therefore has equal magnitude but opposite sign for the two CP asymmetries.

From the general relation 
$\lambda_S + \lambda_L = \lambda 2(a-b) /( 1- b^2 ) $
we learn that a necessary and sufficient condition for $\lambda_S \ne \lambda_L $ to be satisfied
is the presence of non-vanishing wrong-sign amplitudes with $a \ne b$.  
As an example, if each of the right-sign and wrong-sign amplitudes is dominated by a single 
contribution, then $|a| \approx |b|$.  If in addition,
${\rm Re} \lambda \sim {\rm Im} \lambda \sim {\cal O}(1)$, as in
the Standard Model, and ${\rm Arg}[a] \sim {\rm Arg} 
[b] \sim {\cal O}(1)$, we obtain ${\rm Im} (\lambda_S +
\lambda_L ) \sim |a| \sim |b|$.  
Thus the difference in CP asymmetries is of the order of the ratio of wrong-sign to
right-sign kaon amplitudes.

\subsection{Constraints on New Physics Scenarios}

Consider the wrong-sign decay $B^0 \to J/\Psi \overline{K}$.  As the final state does not
contain a $d$ quark, the decay must proceed via annihilation of the $B$ meson.  
It must therefore be mediated by six-quark operators, with an effective 
Hamiltonian of the form
\beq
{g \over M^5 }  b s c \bar c \bar d \bar d \,,
\label{eq:hamiltonian}
\eeq
where $g$ is a dimensionless coupling, $M$ is the scale of new physics ( 
the color indices and Dirac structure of the operators have been suppressed).
An estimate of the wrong-sign amplitude in the factorization approximation yields
\beq 
A_{\bar K}  \sim  {g \over M^5 } f_B f_K f_\Psi m_B m_\Psi (\epsilon_\Psi \cdot p_K )\,.
\label{eq:AKfact}
\eeq
For purposes of comparison, we note that in the Standard Model 
the right-sign amplitude in the factorization approximation is given by 
\beq
A_{K} = {G_F \over \sqrt{2} } V_{cb} a_2 f_\Psi F_1 m_\Psi (\epsilon_\Psi \cdot p_K )\,,
\label{eq:SMAKfact}
\eeq
where $F_1$ is the $B \to K$ form factor, and $a_2$ is a function of the 
current-current operator Wilson coefficients $C_1 $ and $C_2$ arising from $W$ exchange.
The observed $B \to J/\Psi K_S $ amplitude is reproduced if $a_2 \approx .25$.

We can get an order of magnitude upper bound on the quantity $g /M^2 $ by integrating
out the charm quarks at one-loop, yielding a contribution 
to the effective Hamiltonian mediating charmless hadronic $B$ decays of order
\beq
{1 \over 16 \pi^2 } { g\over M^2 } b s \bar d \bar d \,.
\nonumber
\eeq
Upper bounds on the strenghths of such operators can be obtained by considering their contributions 
to the rare decays $B^\pm \to \pi^\pm K_S $ or, more specifically, to the wrong-sign
kaon decays $B^\pm \to \pi^\pm \bar{K}^0 $ \cite{trojan} in the factorization approximation, yielding
$ g/ M^2 \ltap  10^{-5}~{\rm GeV}^{-2}$.  Inserting this bound 
into our estimate for $A_{\bar K}$ 
in Eq.~(\ref{eq:AKfact}) and equating the result 
with the expression
for $A_K$ in Eq.~(\ref{eq:SMAKfact}) (for $a_2 \approx .25$) 
implies that a scale of new physics $M$ of a few
GeV is required in order to obtain significant wrong-sign kaon amplitudes. 

We know of only one scenario with such a potentially low
scale for new flavor-changing interactions: 
supersymmetric models with a light bottom squark $\tilde{b}$ of mass 2-5.5
GeV and light gluinos of mass 12-16 GeV \cite{berger}, which have been proposed
to enhance the $b$ quark production cross section at hadron colliders.
Among the new operators which can arise 
at tree-level are several of the form $\bar d b \tilde{b}^* \tilde{b} $.  Stringent upper bounds on 
their strenghths from rare $B$ decays have been obtained in \cite{sbotfcnc}.
Interactions of the desired form in Eq. (\ref{eq:hamiltonian}) 
would be generated from these operators if 
the $R$-parity violating Yukawa
couplings mediating
$\tilde{b} \to \bar c \bar d$ and $\tilde{b} \to \bar c \bar s$ decays were also present.
Unfortunately, an upper bound of order $10^{-5}$ on the product of these two couplings from 
box-graph contributions to $K - \bar K$
mixing implies that the wrong-sign kaon amplitudes
would be negligibly small, i.e., $a \sim b \sim 10^{-5}$.
Although this result does not constitute a no-go theorem, it appears that
the possibility of significantly different CP asymmetries in $B \to J/\Psi K_{S,L} $ 
decays is extremely unlikely due to the requirement of 
such a low scale for new interactions.

\section{Isospin Violation in $B \to K^* \gamma $}

In the Standard Model, the effective weak Hamiltonian for $b\to s\gamma$
transitions is
\beq
   {\cal H}_{\rm eff} = \frac{G_F}{\sqrt 2} \sum_{p=u,c} \lambda_p^{(s)}
   \bigg( C_1\,Q_1^p + C_2\,Q_2^p + \!\sum_{i=3,\dots,8}\! C_i\,Q_i 
   \bigg)\, ,
\nonumber \eeq
where $\lambda_p^{(s)}=V_{ps}^* V_{pb}$, $Q_{1,2}^p$ are the 
current--current operators, $Q_{3,\dots,6}$ 
are the local four-quark QCD penguin operators, and $Q_7$ and $Q_8$ are the 
electro-magnetic and chromo-magnetic dipole operators. (We adopt the 
conventions of~\cite{BBNS}; in particular, $C_1 \approx 1$ is the largest 
coefficient.)  The Wilson coefficients $C_i$ and the matrix elements of 
the renormalized operators $Q_i$ depend on the renormalization scale 
$\mu$. 

At leading power in $\Lambda/m_b$ the $B\to K^* \gamma$ decay amplitude 
is given by 
\beq
   i{\cal A}_{\rm lead} = \frac{G_F}{\sqrt 2}\,\lambda_c^{(s)} a_7^c\,
   \langle\bar K^*(k,\eta)\gamma(q,\epsilon)|Q_7|\bar B\rangle \,,
\eeq
where the next-to-leading order (NLO) result for the coefficient 
$a_7^c=C_7+\dots$ can be found in \cite{Bosch:2001gv}.  The ellipses denote 
${\cal O}(\alpha_s )$ hard spectator interaction contributions.

\begin{figure}
 \makebox{\includegraphics[height=.2\textheight]{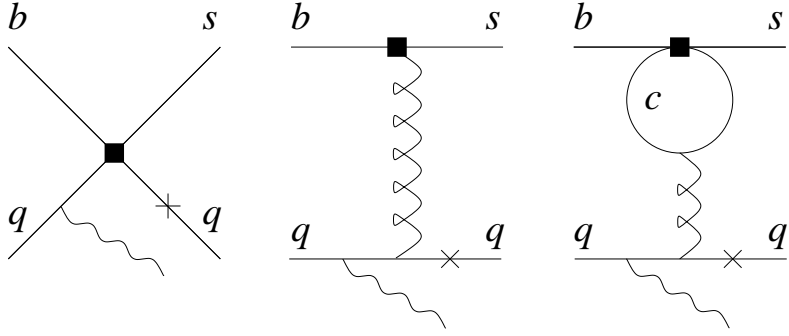}}
  \caption{Spectator-dependent contributions from local 4-quark operators (left), 
the chromo-magnetic dipole operator (center), and the charm penguin 
(right). Crosses denote alternative photon attachments.}
\label{fig:graphs}
\end{figure}

The 
leading isospin-breaking effects arise from the diagrams shown in 
Fig.~\ref{fig:graphs}. 
Their contributions to the decay amplitudes can be parametrized as 
${\cal A}_q=b_q\,{\cal A}_{\rm lead}$, where $q$ is 
the flavor of the spectator antiquark in the $\bar B$ meson. 
We neglect NLO
terms of order $\alpha_s\,C_{3,\dots,6}$ while retaining terms of order 
$\alpha_s\,C_{1,8}$. This is justified, because the penguin coefficients 
$C_{3,\dots,6}$ are numerically very small. Also, it is a safe 
approximation to neglect terms of order 
$\alpha_s\,\lambda_u^{(s)}/\lambda_c^{(s)}$. It then suffices to 
evaluate the contributions of the 4-quark operators shown in the first 
diagram at tree level.  

The QCD 
factorization approach gives an expression for the coefficients $b_q$ 
in terms of convolutions of hard-scattering kernels with light-cone 
distribution amplitudes for the $K^*$ and $B$ mesons. 
The result can be written as 
\beq
   b_q = \frac{12\pi^2 f_B\,Q_q}{m_b\,T_1^{B\to K^*} a_7^c}
   \left( \frac{f_{K^*}^\perp}{m_b}\,K_1 
   + \frac{f_{K^*} m_{K^*}}{6\lambda_B m_B}\,K_2 \right)\,,
\label{eq:bq}
\eeq
where $T_1^{B\to K^*}$ is a form factor in the decomposition of the
$B\to K^*$ matrix element of the tensor current, and $m_b$ denotes the running $b$-quark mass.  
$\lambda_B $ is a hadronic parameter which enters the first inverse moment
of the relevant $B$ meson distribution amplitude.
Heavy-quark scaling laws 
imply that $b_q$
scales like 
$\Lambda/m_b$.  However, because of 
the large numerical factor $12\pi^2$ in the numerator the values of 
$b_q$ turn out to be larger than anticipated in~ 
\cite{Bosch:2001gv,Beneke:2001at}.

The dimensionless quantities $K_1 $ and $K_2 $
are linear functions of the Wilson coefficients.  The latter
are multiplied by convolution integrals of hard scattering kernels with meson distribution
amplitudes.  The integrals associated with the four-quark QCD penguin operators 
and current-current operators 
exist for any reasonable choice of the distribution amplitudes, which shows that
the QCD factorization approach holds at subleading power for their matrix elements,
to the order we are working.  However, the convolution integral associated with the chromomagnetic dipole
operator suffers from a logarithmic end point singularity,
indicating that at 
subleading power factorization breaks down for this matrix element.  
We regulate the singularity by introducing a cutoff.  A large 
uncertainty is assigned to this estimate since this contribution must be dominated 
by soft physics.

To leading-order in small quantities 
the theoretical expression for the isospin-breaking parameter (see
Eq.~(\ref{eq:isospinavg})) is 
$\Delta_{0-}={\rm{Re}}(b_d-b_u)$.
A dominant uncertainty in the prediction for $\Delta_{0-}$ comes from the 
tensor form factor $T_1^{B\to K^*}$, recent estimates of which range 
from $0.32_{\,-\,0.02}^{\,+\,0.04}$~\cite{DelDebbio:1998kr} to 
$0.38\pm 0.06$~\cite{Ball:1998sk}. On the other hand, a fit to the 
$B\to K^*\gamma$ branching fractions yields the lower value 
$0.27\pm 0.04$~\cite{Beneke:2001at}. To good approximation the result 
for $\Delta_{0-}$ is inversely proportional to the value of the form 
factor. We take $T_1^{B\to K^*}=0.3$ (at $\mu=m_b$) as a reference 
value.  Values for the remaining input parameters together with their 
uncertainties can be found in Ref.~\cite{kaganneubert}.

Combining all sources of uncertainty we obtain the Standard Model result
\beq
   \Delta_{0-} = (8.0_{\,-\,3.2}^{\,+\,2.1})\% \times
   \frac{0.3}{T_1^{B\to K^*}} \,.
\nonumber
\eeq
The three largest contributions to the error from input parameter 
variations are due to $\lambda_B$ (${}_{\,-\,2.5}^{\,+\,1.0}\%$), the 
divergent integral in the matrix element of $Q_8$ ($\pm 1.2\%$), and the decay constant $f_B$ 
($\pm 0.8\%$). The perturbative uncertainty is about $\pm 1\%$. Our 
result is in good agreement with the current central experimental value 
of $\Delta_{0-}$ including its sign, which is predicted unambiguously. 
By far the most important source of isospin-breaking is due to the 
four-quark penguin operator $Q_6$, whose contribution to $\Delta_{0-}$ is 
about 9\% (at $\mu \approx m_b$). The other terms are much smaller. In 
particular, the contribution of the chromo-magnetic dipole operator, for 
which factorization does not hold, is less than 1\% in magnitude and 
therefore numerically insignificant. Hence, the important 
isospin-breaking contributions can be reliably calculated using QCD 
factorization. It follows from our result that these effects {\it mainly test 
the magnitude and sign of the ratio ${\rm Re}( C_6/C_7 )$ of penguin coefficients.} 

Because of their relation to matrix elements of penguin operators, 
isospin-breaking effects in $B\to K^* \gamma$ decays are sensitive probes 
of physics beyond the Standard Model. In particular, scenarios in which 
the sign of $\Delta_{0-}$ is flipped could be ruled out in the near future with more 
precise data.  For simplicity lets restrict ourselves to 
new physics models which do not enlarge the Standard Model operator basis,
where this possibility corresponds to flipping the sign of ${\rm Re}( C_6/C_7 )$.
As a specific example, consider the minimal supersymmetric standard model (MSSM)
with minimal flavor violation, and with $tan \beta $
enhanced contributions to $B \to X_s \gamma $ decays taken into account 
beyond leading-order \cite{degrassi}.  
In this scenario new contributions to $Q_{3,\dots,6}$, and $Q_8$ 
are too small to have a significant effect.
For low $\tan \beta $, ${\rm Re} C_7 (m_b )$ is negative as in the Standard Model.
However, for $\tan \beta \gtap 27 $
${\rm Re} C_7 (m_b )$ can take on both positive or negative values with positive values becoming
more probable as 
$tan \beta $ increases~\cite{bozpak}.  Therefore, isospin breaking could rule out
significant regions
of MSSM parameter space at large $tan\beta $.
The sign of ${\rm Re} C_7 $ can also be flipped in supersymmetric models
with non-minimal flavor violation (independently of $tan \beta$) via gluino/down-type squark 
loop graphs, without affecting the sign of $C_6$.  A more detailed 
treatment of new physics effects, including the possibility of an enlarged 
operator basis will be presented elsewhere.

\section{Constraints on the $g^* g \,\eta^\prime $ Form Factor
from $\Upsilon (1 S) $ Decays}

The effective $\eta^\prime g^* g $ coupling can be written as 
\begin{equation}
H(q^2 ) \epsilon_{\alpha \beta \mu \nu}
q^\alpha k^\beta \epsilon_1^\mu \epsilon_2^\nu ,
\label{eq:anomaly}
\end{equation}
where $q = p_b - p_s $ is the virtual gluon's momentum, $k$ is the 
`on-shell{'} gluon's momentum, and 
$H (q^2  )$ is the 
$g^* g \,\eta^\prime $ transition form factor.    
The QCD axial anomaly determines the 
form factor at small momentum transfer, i.e., in the 
$q^2 \to 0$ limit.  An estimate of this limit from the decay rate for
$J/\psi \to \eta^\prime \gamma$ gives~\cite{atwoodsoni}
$H (0) \approx 1.8$ $\rm{GeV}^{-1}$. 

As we have discussed, the crucial issue that needs to be addressed in 
determining the $b \to s g \eta^\prime$
decay rate is the dependence of $H$ on $q^2$. 
A simple model for the $g g \eta^\prime $ vertex \cite{kaganpetrov} 
in which a pseudoscalar current is coupled 
perturbatively to two gluons through
quark loops yields a form factor which can be parametrized as
\beq H (q^2 ) = \frac{H_0  m_{\eta^\prime}^2}{q^2 -m_{\eta^\prime}^2}.
\label{eq:formfactor}
\eeq
The dependence of $H_0$ on $q^2$ is subleading, but it 
insures the absence of a pole at $q^2 = m_{\eta^\prime}^2 $. 
To first approximation it can be modeled by a constant
to be identified with the
low energy coupling extracted from $J/\psi \to \eta^\prime \gamma$, e.g.,
$H_0 \approx 1.8$ $\rm{GeV}^{-1}$.
The above parametrization agrees well with 
recent pQCD calculations of the form factor \cite{muta,aliformfactor}
in which hard amplitudes involving quark and gluon exchanges are convoluted
with the $\eta^\prime$ quark and gluon wave functions, particularly
if $H_0 \approx 1.7$ GeV$^{-1}$.

\begin{figure}
 \makebox{\includegraphics[height=.2\textheight]{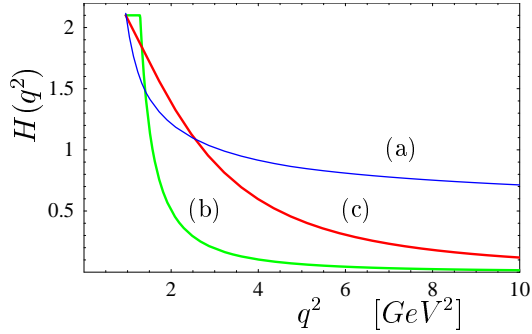}}
  \caption{Three choices for the form factor $H (q^2 )$, as described in the text:
(a) the slowly falling form factor, (b) a rapidly falling
form factor representative of perturbative QCD calculations,
(c) an intermediate example.}
\label{fig:q2dependence}
\end{figure}

Below we will consider three representative choices for
$H(q^2 )$:
\begin{description} 
\item[a)] The slowly falling form factor of Ref.~\cite{houtseng}, 
$H (q^2 ) = \sqrt{N_f} \alpha_s (q^2 ) \cos\theta /(\pi f_{\eta^\prime} )
\approx 2.1 \alpha_s (q^2) /\alpha_s (m_{\eta^\prime}^2 )$ GeV$^{-1}$ ($\theta $ is
the pseudoscalar mixing angle).  It gives ${\rm BR}(b \to sg
\eta^\prime ) \approx 6.8 \times 10^{-4}$ for $p_{\eta^\prime} > 2$ GeV.
\item[b)] The rapidly
falling form factor of Eq.~(\ref{eq:formfactor}), with $H_0 \approx 1.7$ 
$\rm{GeV}^{-1}$ (at low
$q^2 $ it is matched onto
the value of the previous form factor at $q^2 = m_{\eta^\prime}^2 $). It gives
${\rm BR}(b \to sg
\eta^\prime ) \approx 3 \times 10^{-5}$ for $p_{\eta^\prime} > 2$ GeV.
\item[c)] An intermediate purely phenomenological form factor 
$H (q^2 ) \propto 1/(q^2 + M^2 )$ with $M = 2.2$ GeV,
which gives
${\rm BR}(b \to sg \eta^\prime )\approx 4.4 \times 10^{-4}$ for $p_{\eta^\prime} > 2$ GeV.  
\end{description}
The three form factors are plotted in
Fig.~\ref{fig:q2dependence}.  
 
The $g^* g \,\eta^\prime $ form factor 
induces the decay $^3 S_1^{[1]}   \to g g (g^* \to \eta^\prime g )$ of the dominant color-singlet
$\Upsilon (1S) $ Fock state.   
Moreover, the region of
$q^2$ relevant for fast $\eta^\prime $ production in this decay  
has large overlap with the region of $q^2 $ relevant for
fast
$\eta^\prime$ production in $b \to  sg \eta^\prime $ decays.
Therefore, the $\eta^\prime $ spectrum in $\Upsilon (1S)$ decays
could potentially constrain the $g^* g \,\eta^\prime $ form factor, 
and at the same time tell us if the subprocess $b \to sg \eta^\prime$
can account for the $\eta^\prime $ yield in $B$ decays~ \cite{chenkagan}.
  
The ARGUS collaboration mesured inclusive $\eta^\prime $ production
at the $\Upsilon (1S)$~\cite{ARGUS}.  Unfortunately, due to limited
it was not possible to subtract the continuum contribution from the
$\eta^\prime$ spectrum, so that modeling of the direct $\Upsilon (1 S)$ decay 
and continuum components was required.  The term `direct{'} refers to hadronic
$\Upsilon (1S)$ decays with $e^+ e^- \to \Upsilon (1S) \to q \bar q$ 
events subtracted (the latter are included in the continuum data sample), 
essentially leaving $\Upsilon (1S) \to ggg$ events. 
The ARGUS fit for direct decays gives
$d n / dz  \approx 16.8 \beta e^{-10.6 z}$, where
$n$ is the $\eta^\prime $ multiplicity,
and 
$z\equiv 2 E_{\eta^\prime} / m_{\Upsilon} $.
However, it is possible to obtain a model-independent upper bound
on the $\eta^\prime $ multiplicity in each $z$ bin 
by assuming that the raw $\eta^\prime $ yield is entirely due to
direct $\Upsilon (1S)$ decays (note that the total number of direct $\Upsilon
(1S)$ decays in the data sample is known).\footnote{I am grateful to Axel Lindner and 
Dietrich Wegener for supplying the 
raw $\eta^\prime$ yields and efficiencies from 
Ref. \cite{diplomthesis}.}  Of particular interest is the bound for the highest energy 
bin ($.7 < z < 1.0 $) for which we obtain 
$n_{z > .7} < (6.5 \pm 1.3) \times 10^{-4}$, where the error is statistical only.
The model-dependent fit gives a slightly larger result, $8.8 \times 10^{-4}$.  
We note here that in the near future the CLEO and CLEO-C collaborations will be able to
measure the $\eta^\prime $ spectrum in $\Upsilon (1S)$ decays
far more precisely.

\begin{figure}
 \makebox{\includegraphics[height=.2\textheight]{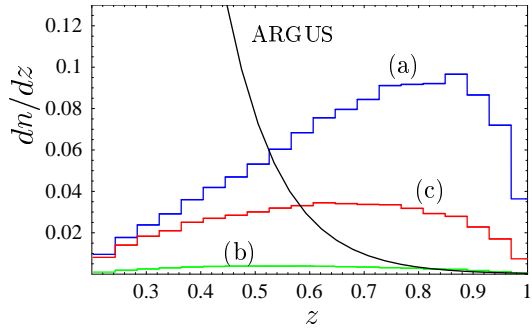}}
  \caption{The $\eta^\prime $ multiplicity spectrum $d n/dz$ in direct $\Upsilon (1S) $ decays 
for the three $g^* g \,\eta^\prime $ form factors (a) Slowly falling, (b) pQCD
(c) intermediate, and the model-dependent ARGUS fit, as described in the text.}
\label{fig:spectra}
\end{figure}

We have calculated the $\eta^\prime $ spectrum, $d \Gamma (^3 S_1^{[1]}   \to g g
g \eta^\prime ) /dz $ for each of the three $g^* g
\,\eta^\prime$ form factors considered above at leading-order in 
QCD, using a four-body phase space Monte
Carlo \cite{chenkagan}. Relativistic corrections
have not been taken into account.  
The spectra have been normalized with respect to the 
leading-order three gluon decay width, $\Gamma (^3 S_1^{[1]}   \to g g g )$,
to obtain the leading-order $\eta^\prime $ multiplicities.
The results are compared to the ARGUS fit in 
Fig.~\ref{fig:spectra}.  
Furthermore, comparison with the 
model-independent ARGUS upper bound for $z> .7$ gives
\begin{equation}
  \left({n^{ \rm theory}  \over
n^{ \rm argus}}\! \right)_{\! z>.7}
\gtap \cases{41  & {\rm (a) slowly falling}, \vspace{0.1cm} \cr
   1  & {\rm (b) pQCD}, \vspace{0.1cm} \cr
        13   & {\rm (c) intermediate}. \cr}
\label{eq:BRratios}
\end{equation}
Evidently, only the rapidly falling (pQCD) form factor 
is consistent with the ARGUS data at large energies.  The intermediate and slowly falling form factors 
are in gross conflict, giving order of magnitude or greater excesses.
Note that the bulk of the $\eta^\prime $ yield is expected to originate from long-distance fragmentation
of the three gluon configuration and should therefore be quite soft, in agreement with the ARGUS data.

We have checked that significant contributions from 
decays of higher Fock states of the $\Upsilon (1S) $ induced by the
$g^* g \,\eta^\prime $ coupling, e.g., 
$^1 S_0^{[8]}, ^3 P_J^{[8]} \to gg \eta^\prime $, would only harden
the $\eta^\prime $ spectra. 
Furthermore, relativistic corrections, next-to-leading order QCD
corrections, and other theoretical refinements would be insufficient to eliminate 
the large excesses for $z>.7$ in Eq.~(\ref{eq:BRratios}). Finally, we find that even in the extreme 
case that the gluon is given a `phenomenological{'} 1 GeV mass, 
as advocated in Ref. \cite{field} (to
improve agreement between the predicted and
observed photon spectra near the end-point for $\Upsilon (1S)$ decays), the slowly falling
form factor still gives an order of magnitude excess.

\section{Conclusion}

We have seen that a necessary condition for obtaining 
significantly different time-dependent CP asymmetries in $B \to J/\Psi K_{S,L} $
is a non-vanishing
amplitude for the `wrong sign{'} kaon decay
$B \to J/\Psi \overline{K}$ or its CP conjugate. 
We note that it may be possible to test directly for the presence of these amplitudes
by searching for wrong-sign $B^0 \to J/\Psi (\bar{K}^*  \to K^+ \pi^- )$ decays.
A model-independent analysis shows that the scale of new physics needs to be prohibitively low,
of order a few GeV, all but ruling out this possibility. 
The only potential example we have found,
in the framework of supersymmetric models with an ultra-light bottom squark and light gluinos,
would lead to gross violation of $K-\bar K$ mixing constraints.

We have seen that isospin breaking effects in $B \to K^* \gamma $ decays 
can be calculated model-indepndently in the QCD factorization framework.
In the Standard Model the decay rate for 
$\bar B^0\to\bar K^{*0}\gamma$ is predicted to be about 10--20\% larger 
than that for $B^- \to K^{*-}\gamma$, in agreement with the measured central values.
The direction of the inequality between these amplitudes is particularly sensitive to the sign 
of the Wilson coefficient ratio ${\rm Re} (C_6 / C_7 )$.  As an application, more precise data
which will be available in the not too-distant future
could rule out models in which the sign of ${\rm Re} C_7 $ is flipped relative to the 
Standard Model.

Finally, we have seen that slowly falling $g^* g \,\eta^\prime $ form factors which can 
explain the large $B \to \eta^\prime X_s $ rate via the subprocess
$b \to s (g^* \to g \eta^\prime ) $ are ruled out by ARGUS data on
fast $\eta^\prime $ production in 
$\Upsilon (1S ) $ decays.  However, the rapidly falling form 
factor predicted by perturbative QCD is compatible.
Unfortunately, the corresponding $B \to \eta^\prime X_s $ rate for $p_{\eta^\prime } > 2$ GeV 
is a factor of 20 smaller than 
observed.
An enhanced $b \to s g$ chromo-magnetic dipole operator, 
motivated by the 
low semileptonic branching ratio and charm multiplicity in $B$ decays, 
could improve agreement with experiment \cite{kaganpetrov}.
The CLEO and CLEO-C collaborations will be able to make much more precise measurements 
of the $\eta^\prime $ spectrum in $\Upsilon (1S)$ decays in the near future, 
potentially providing even more stringent constraints on the $g^* g \,\eta^\prime$
form factor.

\begin{theacknowledgments}
It is a pleasure to thank my collaborators on the works presented here, 
Yixiong Chen, Yuval Grossman, 
Matthias Neubert, and Zoltan Ligeti.
I am grateful to the organizers for arranging a very stimulating symposium, 
under very trying circumstances.  I would also like to thank Axel Lindner, Alexey Petrov, Soeren Prell,
Zack Sullivan, and Dieterich Wegener for useful discussions.
\end{theacknowledgments}

\end{document}